\documentclass[authoryear, 5p, number]{elsarticle}

\usepackage{amsmath}
\usepackage{txfonts}
\usepackage{booktabs}
\usepackage{array}
\usepackage{graphicx}
\usepackage{natbib,twoopt}
\usepackage{upgreek}
\usepackage{color}
\usepackage[dvipsnames]{xcolor}
\usepackage[breaklinks=true]{hyperref} 
\usepackage{lineno,hyperref}
\usepackage{subcaption}
\modulolinenumbers[5]

\journal{Astronomy \& Computing}

\begin{document}
	
	\begin{frontmatter}
		
		\title{Apercal - The Apertif Calibration Pipeline}
		
		\author[a,b]{B.~Adebahr\corref{cor1}}
		\ead{adebahr@astro.rub.de}
		
		\author[b]{R.~Schulz}
		\author[b]{T.~J.~Dijkema}
		\author[c,d,b]{V.~A.~Moss}
		\author[b,f]{A.~R.~Offringa}
		\author[b,h]{A.~Kutkin}
		\author[f]{J.~M.~van~der~Hulst}
		\author[i,j,k]{B.~S.~Frank}
		\author{N.~P.~E.~Vilchez}
		\author[f,b]{J.~Verstappen}
		\author[b,f]{E.~K.~Adams}
		\author[b,k,f]{W.~J.~G.~de~Blok}
		\author[b]{H.~Denes}
		\author[b,f]{K.~M.~Hess}
		\author[g]{D.~Lucero}
		\author[b,f]{R.~Morganti}
		\author[b,f]{T.~Oosterloo}
		\author[l]{D.-J.~Pisano}
		\author[m]{M.~V.~Ivashina}
		\author[b]{W.~A.~van~Cappellen}
		\author[n,b]{L.~D.~Connor}
		\author[b]{A.~H.~W.~M.~Coolen}
		\author[b]{S.~Damstra}
		\author[b]{G.~M.~Loose}
		\author[b]{Y.~Maan}
		\author[q]{F.~M.~Maccagni}
		\author[b]{A.~Mika}
		\author[b]{H.~Mulder}
		\author[b,n,p]{L.~C.~Oostrum}
		\author[b]{E.~Orr\'u}
		\author[b]{R.~Smits}
		\author[b]{D.~van~der~Schuur}
		\author[b,n]{J.~van~Leeuwen}
		\author[b]{D.~Vohl}
		\author[b]{S.~J.~Wijnholds}
		\author[b,o]{J.~Ziemke}

		\cortext[cor1]{Corresponding author}
		
		\address[a]{Astronomisches Institut der Ruhr Universit\"at Bochum (AIRUB), Bochum, Germany}
		\address[b]{ASTRON, Dwingeloo, The Netherlands}
		\address[c]{CSIRO Astronomy and Space Science, Sydney, Australia}
		\address[d]{University of Sydney, Sydney, Australia}
		\address[e]{Cavendish Laboratory, Cambridge, UK}
		\address[f]{Kapteyn Astronomical Institute, University of Groningen, Groningen, The Netherlands}
		\address[g]{Department of Physics, Virginia Tech, Blacksburg, VA, USA}
		\address[h]{Astro Space Center of Lebedev Physical Institute, Profsoyuznaya Str. 84/32, 117997 Moscow, Russia}
		\address[i]{South African Radio Astronomy Observatory (SARAO), 2 Fir Street, Observatory, 7925, South Africa}
		\address[j]{Inter-University Institute for Data Intensive Astronomy and Department of Astronomy, University of Cape Town, Private Bag X3, Rondebosch 7701, South Africa}
		\address[k]{University of Cape Town, Cape Town, University Private Bag Rondebosch, 7701, South Africa}
		\address[l]{Department of Physics and Astronomy, West Virginia University, White Hall, Box 6315, Morgantown, WV 26506, USA}
		\address[m]{Dept.\ of Electrical Engineering, Chalmers University of Technology, Gothenburg, Sweden}
		\address[n]{Anton Pannekoek Institute, University of Amsterdam, Postbus 94249, 1090 GE Amsterdam, The Netherlands}
		\address[o]{Rijksuniversiteit Groningen Center for Information Technology, P.O. Box 11044, 9700 CA Groningen, the Netherlands}
		\address[p]{Netherlands eScience Center, Science Park 140, 1098 XG, Amsterdam, The Netherlands}
		\address[q]{INAF – Osservatorio Astronomico di Cagliari, Via della Scienza 5, 09047 Selargius, CA, Italy}

		\begin{abstract}
			
			Apertif (APERture Tile In Focus) is one of the Square Kilometre Array (SKA) pathfinder facilities. The Apertif project is an upgrade to the 50-year-old Westerbork Synthesis Radio Telescope (WSRT) using phased-array feed technology. The new receivers create 40 individual beams on the sky, achieving an instantaneous sky coverage of 6.5 square degrees. The primary goal of the Apertif Imaging Survey is to perform a wide survey of 3500 square degrees (AWES) and a medium deep survey of 350 square degrees (AMES) of neutral atomic hydrogen (up to a redshift of 0.26), radio continuum emission and polarisation. Each survey pointing yields 4.6\,TB of correlated data. The goal of \emph{Apercal} is to process this data and fully automatically generate science ready data products for the astronomical community while keeping up with the survey observations. We make use of common astronomical software packages in combination with Python based routines and parallelisation. We use an object oriented module-based approach to ensure easy adaptation of the pipeline. A Jupyter notebook based framework allows user interaction and execution of individual modules as well as a full automatic processing of a complete survey observation. If nothing interrupts processing, we are able to reduce a single pointing survey observation on our five node cluster with 24 physical cores and 256\,GB of memory each within 24h keeping up with the speed of the surveys. The quality of the generated images is sufficient for scientific usage for 44\,\% of the recorded data products with single images reaching dynamic ranges of several thousands. Future improvements will increase this percentage to over 80\,\%. Our design allowed development of the pipeline in parallel to the commissioning of the Apertif system.
			
		\end{abstract}
		
		
	\end{frontmatter}

	\section{Introduction}
	
	Radio astronomy is currently moving into a survey era with most of the new or upgraded facilities being designed as survey telescopes. The technological advances over the past few decades in receiver design allow the usage of broadband receivers with fine spectral resolution. In addition, Phased Array Feeds (PAFs) are enlarging the field of view by an order of magnitude by forming tens of adjacent beams simultaneously. The increase in computing power and easy access to powerful computing facilities allows the use of these receiver technologies for radio interferometric imaging. The amount of data produced by modern radio interferometric observations already reaches several terabytes for observation times of a few hours posing a new challenge to the processing of the data.
	
	The Apertif system is an upgrade of the receiver system of the old Westerbork Synthesis Radio Telescope (WSRT) located in the Netherlands with PAF technology. Until January 2021 the receivers were operating at a central L-band frequency of 1280\,MHz, which was moved to 1370\,MHz afterwards. The instantaneous bandwidth is 300\,MHz. 384 individual subbands consisting of 64 spectral channels each provide a spectral resolution of 12.2\,kHz. 40 individual adjacent beams are formed on the sky achieving an instantaneous sky coverage of 6 square degrees. Voltages are correlated to generate data in all four Stokes parameters. These specifications make Apertif an ideal survey instrument for continuum, polarisation and HI-surveys of the Northern sky. For a detailed overview of the Apertif system we refer the reader to \citet{2021arXiv210914234V}.
	
	The deluge of data produced by this new generation of radio telescopes requires an automatic and/or semi-automatic approach. Several calibration pipelines have been designed over the last few years with different purposes in mind. Many aim for general data reduction of a broad range of different imaging observations, mostly data from a specific facility, like e.g. the CASA pipeline for the Jansky Very Large Array (JVLA) and the Atacama Large Millimeter Array \citep{2018AAS...23134214K, 2014ASPC..485..383M}, DP3\footnote{\url{https://github.com/lofar-astron/DP3}} and prefactor\footnote{\url{https://github.com/lofar-astron/prefactor}} for the Low Frequency Array (LOFAR), FLAGCAL for the Giant Metrewave Radio Telescope (GMRT) \citep{2012ExA....33..157P}, ASKAPsoft\footnote{\url{https://www.atnf.csiro.au/computing/software/askapsoft/sdp/docs/current/pipelines/introduction.html}} for the Australian Square Kilometre Array Pathfinder (ASKAP), CA\-RACal\footnote{\url{https://github.com/caracal-pipeline/caracal}} (formerly MeerKATHI) \citep{2020arXiv200602955J,2020ascl.soft06014J} for the Meer\-KAT telescope\footnote{For some recent publications CARACal was also used for reducing VLA and GMRT data} or rPICARD for Very Long Baseline Interferometry \citep{2019arXiv190501905J}. Others are tailored towards individual surveys like the VLA Sky Survey (VLASS) pipeline \citep{2019AAS...23344602M} or on a specific aspect like e.g. the improvement of the image quality of LOFAR data using a direction dependent calibration technique \citep{2016ApJS..223....2V, 2018A&A...611A..87T} or even completely new approaches like sto\-chastical imaging \citep{2020MNRAS.493.6071Y}. An example for the inclusion of older software packages into a modern pipeline is the Obit-based\footnote{\url{https://www.cv.nrao.edu/~bcotton/Obit.html}} MeerKAT SDP pipeline\footnote{\url{https://github.com/ska-sa/katsdpcontim}}. First steps have been undertaken to use supercomputing architecture for these pipelines \citep{2019A&C....2800293M, 2017A&C....19...75S, 2019arXiv191013835D}. The goal of Apercal, the dedicated imaging and calibration pipe\-line for the Apertif system is to provide an easy to use framework generating science ready images to the user in an automatic way while keeping modularity and transparency.
	
	Apertif conducts an imaging and a time-domain survey \citep{2019NatAs...3..188A}. Here we focus on the imaging survey which comprises two tiers: The Apertif Wide Extragalctic Survey (AWES) and the Apertif Medium deep Extragalactic Survey (AMES) which are designed to cover 3500 and 350 square degrees, respectively. These surveys are able to probe the neutral hydrogen content of the Universe up to a redshift of 0.26 and the faint general and polarised radio source population. The footprints of the surveys are vastly overlapping with the ongoing LOFAR LoTSS \citep{2017A&A...598A.104S} and future WEAVE \citep{2016sf2a.conf..271S} and LOFAR LBA Sky Survey (LoLSS) \citep{2021arXiv210209238D}. In addition, the survey areas have been chosen such that the observing efficiency is optimised and that there is good overlap with regions that have important ancillary photometric and spectroscopic data in the optical and the IR. SDSS\footnote{\url{www.sdss.org}} \citep{2017AJ....154...28B}, Pan-STARRS\footnote{\url{panstarrs.ifa.hawaii.edu/pswww/}} \citep{2016arXiv161205560C}, and WISE\footnote{\url{https://wise2.ipac.caltech.edu/docs/release/allwise/}} \citep{2010AJ....140.1868W} provide optical and near IR photometric coverage over large portions of the surveys while Herschel-Atlas \footnote{\url{www.h-atlas.org}} \citep{2016MNRAS.462.3146V} provides far IR photometric coverage over a portion of AMES. ManGA\footnote{\url{www.sdss.org/surveys/manga/}} \citep{2016AJ....152..197Y} and WEAVE\footnote{\url{www.ing.iac.es/weave/about.html}}\citep{2014SPIE.9147E..0LD} provide targeted spectroscopic coverage of optically selected galaxies and/or radio selected targets. HetDex\footnote{\url{hetdex.org}}\citep{2008ASPC..399..115H} will provide untargeted spectroscopic coverage over a large area of the northern sky. These synergies allow multi-wavelength studies for a statistically significant number of objects.
	
	The Apertif system is installed on 12 (RT2-RTD) of the 14 dishes of the existing WSRT array \citep{2021arXiv210914234V} providing maximum baselines of 2.4\,km. The array is aligned in an East-West-configuration, which requires nearly full synthesis 12\,hr tracks for sufficient (u,v)-coverage to allow good imaging performance. A single survey pointing produces approximately 4.6\,TB of raw visibility data. This large data volume provides a challenge to the calibration and analysis of the data, making the traditional hands-on approach unfeasible. Over the last decades several data reduction pipelines have been developed using software packages like The Netherlands East West Synthesis Telescope Array Reduction (NEWSTAR)\footnote{\url{https://github.com/lofar-astron/Newstar}}, the Astronomical Image Processing System (AIPS) \citep{Wells1985} and the Multichannel Image Reconstruction Image Analysis and Display (MIRIAD) \citep{1995ASPC...77..433S}.
	
	The similarity of the Apertif data to the old WSRT data, together with the experience we collected from previous WSRT data reduction schemes, allowed us to design \emph{Apercal}, a pipeline which is using modern astronomical software and suits the specific needs of the Apertif system.
	
	In this paper we describe the workflow of \emph{Apercal} including the routines used for reducing the data. The challenges of building a stable software package with exchangeable modules, automatic execution and user-interaction are highlighted. In Sect. \ref{sect_surveyoperations} we give an overview of the Apertif surveys. We describe the design and workflow of the Apercal pipeline in Sect. \ref{sect_apercaldesign} followed by detailed explanations of the individual modules and subroutines in Sect. \ref{sect_pipelinemodules}. An additional \emph{Apercal} independent mosaicking routine is described in Sect. \ref{sect_mosaicking} The final data products and their ingest into the Apertif archive is described in Sect. \ref{sect_finaldata}. The quality assessment of the data products is illustrated in Sect. \ref{sect_dataquality}. We summarise and give an outlook to future improvements of \emph{Apercal} in Sect. \ref{summary}.
	
	\section{Apertif survey operations}
	\label{sect_surveyoperations}
	
	In this section we give a brief overview of how Apertif imaging observations are executed and of completed and ongoing survey operations. For a more detailed description we refer the reader to Adams et al. in prep. and Hess et al. in prep.
	
	Each Apertif imaging observation consists of two different types of observations: Calibrator and target observations. For each of the 40 beams in a target observation usually two calibrator observation are performed. Calibrator observations consist of short scans of 3-5 minutes on a standard reference flux calibration source (3C147 or 3C196) and a standard reference polarised calibration source (3C138 or 3C286). These observations are necessary for an absolute flux, polarisation leakage and polarisation angle calibration and are not used for any imaging. Due to the equatorial mount of the WSRT dishes, which keeps the beam pattern fixed on the sky while tracking, and the good amplitude and phase stability of the Apertif system, these scans only have to be scheduled every 12 to 24\,hrs.
	
	Target survey fields are observed uninterrupted for 11.5\,hrs and are bracketed by two calibrator observations, a flux calibrator and a polarised calibrator source since full polarimetric calibration needs unpolarised as well as polarised calibrator solutions. The 11.5\,hr exposure was found to be the best compromise between sufficient (u,v)-coverage guaranteeing the image quality and sensitivity for spatial scales up to several arcminutes, which is needed for the science goals of the surveys, and the optimal usage of telescope time for executing the survey.
	
	\section{\emph{Apercal} design}
	\label{sect_apercaldesign}
	
	\emph{Apercal} is a radio astronomical data reduction pipeline for Apertif imaging observations which is based entirely on py\-thon2.7 \citep{van1995python}. The pipeline in its final stage will be frozen to this python version. Additional software also incorporating data analysis routines and exchanges of pipeline modules with improvements to the pipeline routines will be developed in python3.
	
	Each individual calibration step in the pipeline is encapsulated into a single module. This approach, by defining the input and output formats of the data and using independent data sub-directories, guarantees the independence of each individual module. It allows the exchange of single routines or even entire pieces of the software with minimal effort and impact onto the rest of the pipeline. Each module consists of individual subroutines, which are handling specific sub-calibration and -imaging tasks within a module. Each subroutine can be imported and executed independently. The whole pipeline is compatible with the Python Jupyter notebook framework\footnote{\url{https://jupyter-notebook.readthedocs.io/en/stable/notebook.html}}, so that user-interaction is simple and pipeline parameters can be changed on-the-fly. During the commissioning of the Apertif system details and parameters of the data reduction still had to be evaluated and were subject to several changes. At this stage the modular \emph{Apercal} structure allowed a minimisation of the workload for the pipeline developers and commissioning team.
	
	\emph{Apercal} configuration parameters are read in by the pipeline via a simple text file. An example is given in \ref{appendix_configfile}. Information in this file consist of the location of the data and the calibration parameters for the individual modules. If a parameter is not given in a file, a default value is read from a master configuration file.
	
	Communication and transfer of parameters between modules is performed via a parameter file inside which the outcome of each individual calibration step is saved and can be accessed. An example can be found in \ref{appendix_paramfile}. Actual logging of information, warning or error messages from the pipeline is carried out via the python logging module. After an execution of a subroutine the parameter file is updated on disk saving the exact parameters used and the success or failure of the task. Entries in the parameter file are organised in a bibliography structure with a string-keyword and an attached data structure of any needed format. Each module and subroutine has certain prerequisites, which are mostly based on the quality and success or failure of a previous calibration step. Prerequisites for the execution of a subroutine or module are checked by loading the parameter file and checking the data values attached to the called bibliography entry. If prerequisites are met, the module is either executed or skipped. This logic is very important for a stable and smooth execution of the pipeline.
	
	The similarity of the reduction strategy and the format of Apertif data to the old WSRT data allowed us to use common astronomical software packages in our pipeline instead of developing our own radio astronomical calibration software. This development philosophy offered the advantage of using reliable and fully tested calibration routines while minimising the workload for software development. Therefore, \emph{Apercal} uses third-party astronomical software, most notably AOFlagger \citep{2012A&A...539A..95O}, the Common Astronomical Software Application (CASA) \citep{2007ASPC..376..127M}, Multichannel Image Reconstruction Image Analysis and Display (Miriad) \citep{1995ASPC...77..433S} and the Python Blob Detector and Source Finder (PyBDSF) \citep{2015ascl.soft02007M}.
	
	Installation of python based modules was performed using the pip-repositories\footnote{\url{https://pypi.org/project/pip/}}. For CASA the available binary installation from the NRAO webpage was used\footnote{\url{https://casa.nrao.edu/}}. PyBDSF was installed via the KERN-repository\footnote{\url{https://kernsuite.info/}}. MIRIAD was compiled with some minor changes (see \ref{appendix_miriad}) using the native ATNF version available on their website\footnote{\url{https://www.atnf.csiro.au/computing/software/miriad/}}. 
	
	Data processing of Apertif data currently takes place on a computing cluster with five individual nodes with identical specifications. Each node is equipped with 24 physical cores with hyperthreading for a total of 48\,CPUs, 256\,GB memory and 33\,TB storage space. Each new observation is first sent to the Apertif Long Term Archive (ALTA) from where it is automatically retrieved and copied to the cluster. The visibilities of each beam of an observation are stored as an individual dataset. This strategy was found to be optimal in terms of storage and processing since individual beams can be and are treated as independent observations. The 40 beams of an observation, with a data size of 115\,GB each, are therefore distributed evenly over four of the nodes leaving the fifth node available for testing, development and further data analysis. Hyperthreading was activated for routines which supported the feature and proved to accelerate the processing.
	
	\emph{Apercal} currently produces data products such as calibration solution tables, fully calibrated target datasets and images in all Stokes parameters as well as line cubes for AWES. Additional processing, e.g. the co-addition of different observing epochs, outside of \emph{Apercal}, is necessary for AMES. Science-ready data products will become public\footnote{\url{https://www.astron.nl/telescopes/wsrt-apertif/apertif-dr1-documentation/}} after a limited propriety time via ALTA\footnote{\url{https://alta.astron.nl/}}. \emph{Apercal} is publicly available on GitHub\footnote{\url{https://github.com/apertif/apercal}}.
	
	\section{Pipeline modules}
	\label{sect_pipelinemodules}
	
\begin{figure*}
		\resizebox{\hsize}{!}{\includegraphics{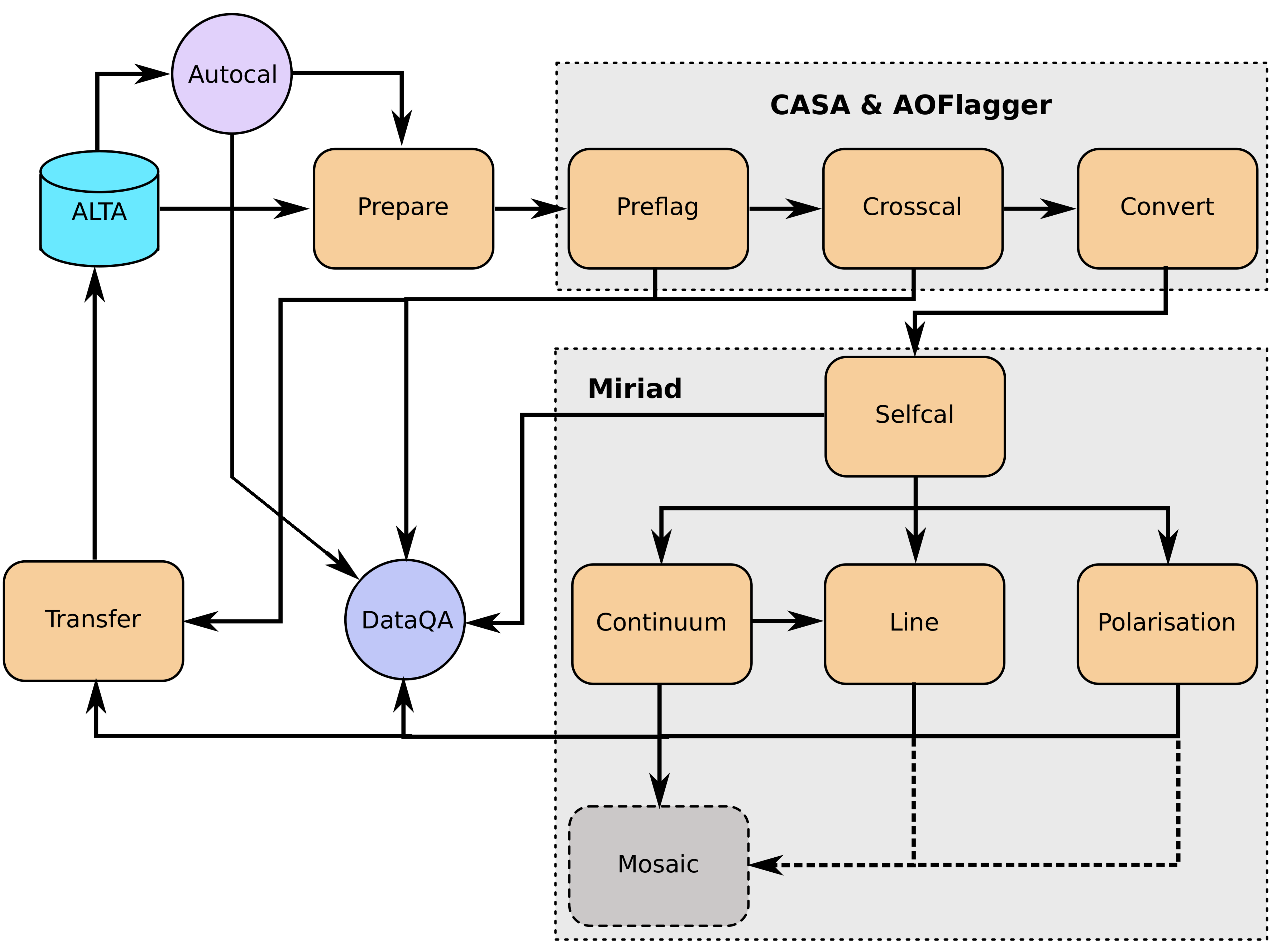}}
		\caption{Overview of the structure of Apercal. Each rectangular box represents a single module. The grey boxes encapsulate the astronomical software packages used within the individual modules. Arrows illustrate the data and workflow within the pipeline. The dashed arrows are routines which are currently developed. The individual modules are described in detail in the following sections: \emph{Autocal}: Sect. \ref{subsect_autocal}, \emph{Prepare}: Sect. \ref{subsect_prepare}, \emph{Preflag}: Sect. \ref{subsect_preflag}, \emph{Crosscal}: Sect. \ref{subsect_crosscal}, \emph{Convert}: Sect. \ref{subsect_convert}, \emph{Selfcal}: Sect. \ref{subsect_selfcal}, \emph{Continuum}: Sect. \ref{subsect_continuum}, \emph{Line}: Sect. \ref{subsect_line}, \emph{Polarisation}: Sect. \ref{subsect_polarisation}, \emph{Transfer}: Sect. \ref{subsect_transfer}, \emph{Mosaic}: Sect. \ref{subsect_mosaic} and Sect. \ref{sect_mosaicking}, \emph{DataQA}: Sect. \ref{sect_dataquality}}
		\label{image_apercaloverview}
	\end{figure*}

\begin{figure}
		\resizebox{\hsize}{!}{\includegraphics{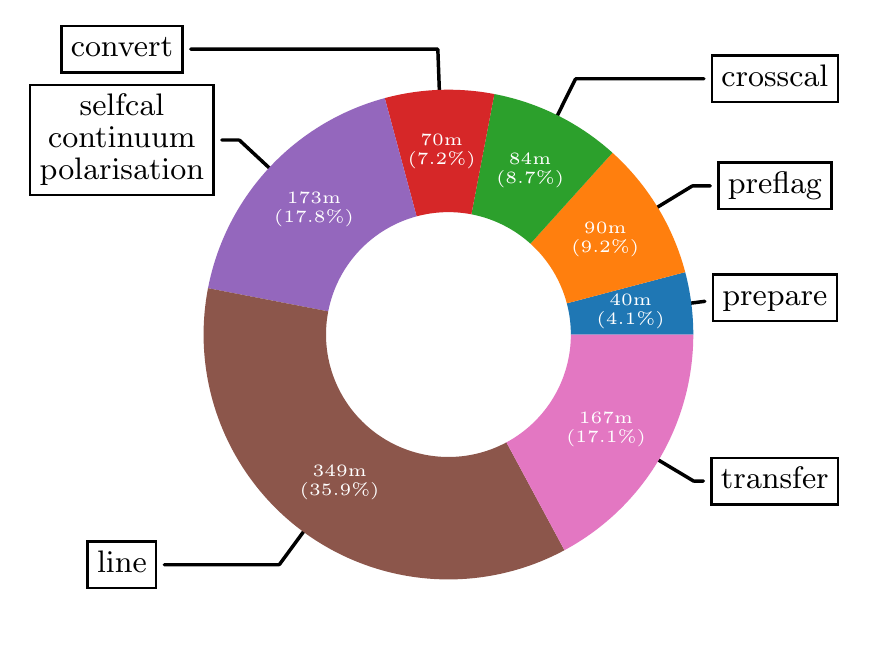}}
		\caption{Pie chart showing the runtimes of the individual \emph{Apercal} modules for an example reduction of one beam. The complete runtime was 972 minutes.}
	    \label{plot_piechart}
	\end{figure}

	In the following section we describe the functionality, the input and output, and the different algorithms used for processing data for each individual pipeline module. An overview of the whole reduction pipeline is given in Fig. \ref{image_apercaloverview}. The whole processing can be divided into four different steps: The triggering of the pipeline involving downloading and preparation of the data, the flagging and cross-calibration, the self-calibration and imaging, and as a last step, the validation and ingest into the Apertif archive. The progress of the pipeline in between and within modules is saved in a stats-file. An overview of the runtime for each individual module is shown in Fig. \ref{plot_piechart}.
	
	\subsection{Autocal}
	\label{subsect_autocal}

	\emph{Autocal} is an automatic routine which executes the start of the reduction pipeline as soon as new observations have been found on ALTA. It automatically identifies the flux and polarisation calibrator data products for a given target field dataset. The necessary information for an \emph{Apercal} pipeline run are then collected and written to a configuration file. After that \emph{Apercal} is triggered and starts downloading the relevant data from ALTA.
	
	Notifications for observers on duty are provided at the start of a pipeline run, in case of failed reduction steps and for successful runs. In case of failed processing steps, the observers are inspecting the data and processing logs for possible reasons. Once the reason has been found and corrected, the failed reduction steps are restarted.
	
	After \emph{Apercal} finishes a reduction \emph{Autocal} automatically triggers the quality assessment (QA) pipeline and ingests the processed data products back into ALTA, with notifications for each stage. Coordinating activities in this way via \emph{Autocal} enables us to moderate the usage of the cluster. 
	
	\subsection{Prepare}
	\label{subsect_prepare}
	
	\emph{Apercal} defines a directory structure for processing where each module uses its own subdirectory to access data and save outputs. All of the following modules (except \emph{convert}) use a single subdirectory, so that individual steps can easily be deleted and restarted. The naming of the directories can be adjusted to the needs of the users with keywords in the configuration file.
	
	The main tasks of the \emph{prepare} module are the setup of the directory structure and the download of the data. Once the module is executed given an input target and calibrator datasets, it checks the availability of the data on the local disc. In case data is not locally available, the module checks the availability on ALTA via an iRODS framework\footnote{\url{https://irods.org/}}. If successful, a Python routine is used\footnote{\url{https://github.com/cosmicpudding/getdata_alta}} to download the data to the local disc and to place it in the appropriate position of the directory structure. After a dataset has been successfully copied from ALTA or located on disc, the correctness of the file is checked via a checksum.
	
	A minimum of a target dataset and a flux calibrator needs to be present per beam for this step to be successful. This condition ensures that, if no flux calibrator is available, the execution of the pipeline is stopped. On the other hand, the pipeline will continue when a polarised calibrator is not available. In this case, the polarisation calibration within the \emph{crosscal} module is omitted and no polarisation imaging is performed. 
	
	\subsection{Preflag}
	\label{subsect_preflag}
	
	The \emph{preflag} module handles all pre-calibration flagging of the data. It is separated into three different operations: The flagging of data with issues known \emph{a priori}, additional flagging options, and fully automatic flagging to identify and mitigate spurious radio frequency interference (RFI). The first two operations use the \textsc{drive-casa} Python wrapper \citep{2015A&C....13...38S} to parse commands to CASA while the last one uses the AOFlagger routines \citep{2012A&A...539A..95O, 2012MNRAS.tmp.2693O, 2010MNRAS.405..155O}.
	
	The subroutines for \emph{a priori} known issues cover three distinct operations: First the data are checked for shadowing. This occurs when the illumination of a dish is hampered by other close-by dishes if a source is observed at a low declination. Timestamps of shadowing can be calculated using a given dish diameter, the position of the dishes and the position of the observed source. Data points positively identified to suffer from shadowing are then automatically flagged. The next step is a mitigation of the effect of the steep bandpass edges of the individual subbands of the Apertif system. The first two and last channels of each 64 channel subband are flagged. Thirdly, due to a finite impulse response (FIR) filter in the channel filterbank, artificial signal is generated in channels 16 and 48 of each subband (see \citet{2021arXiv210914234V}). This causes an artificial source on the order of several $\upmu$Jy in the pointing centre of each image. To mitigate this effect, an option exists to flag these channels automatically.
	
	Subroutines for the additional flagging options encompass the removal of auto correlations, entire antennas, specific cross-correlations, individual baselines, channel and/or time ranges. Any flagging commands not covered by the standard commands can be parsed to a file using the standard CASA-syntax. These options are supposed to be used either when known elements within the Apertif system are not working or a user identifies additional issues during calibration of the data. For the latter case a complete recalibration of the data with the specified additional flags is usually performed. The data ranges to flag for the above mentioned subroutines are specified in the configuration file. More flagging criteria, based on the calibration solutions, are evaluated in the \emph{crosscal} step (see Sect. \ref{subsect_crosscal}).
	
	The last step of the module uses the AOFlagger routines to automatically identify and flag any unknown and previously not flagged RFI in the calibrator and target datasets. Automated detection of interference is challenging because of the large variety in observation properties: calibrators for the surveys are only observed for 3-5 minutes and show high signal-to-noise ratios, whereas the target fields are observed for 11.5\,hours that normally have low signal-to-noise ratios. In the first case, low-level RFI that potentially lasts longer than the calibrator run is hard to identify. In the second case, more sensitive detection is possible, but care has to be taken not to overflag the data. This is especially the case for observations with strong HI-line emission, which could easily be false identified as RFI. We designed a flagging strategy which suits both types of observations and avoids flagging strong line emission.
	
	\begin{figure*}
		\centering
		\begin{subfigure}[c]{0.5\textwidth}
			\includegraphics[width=0.98\columnwidth]{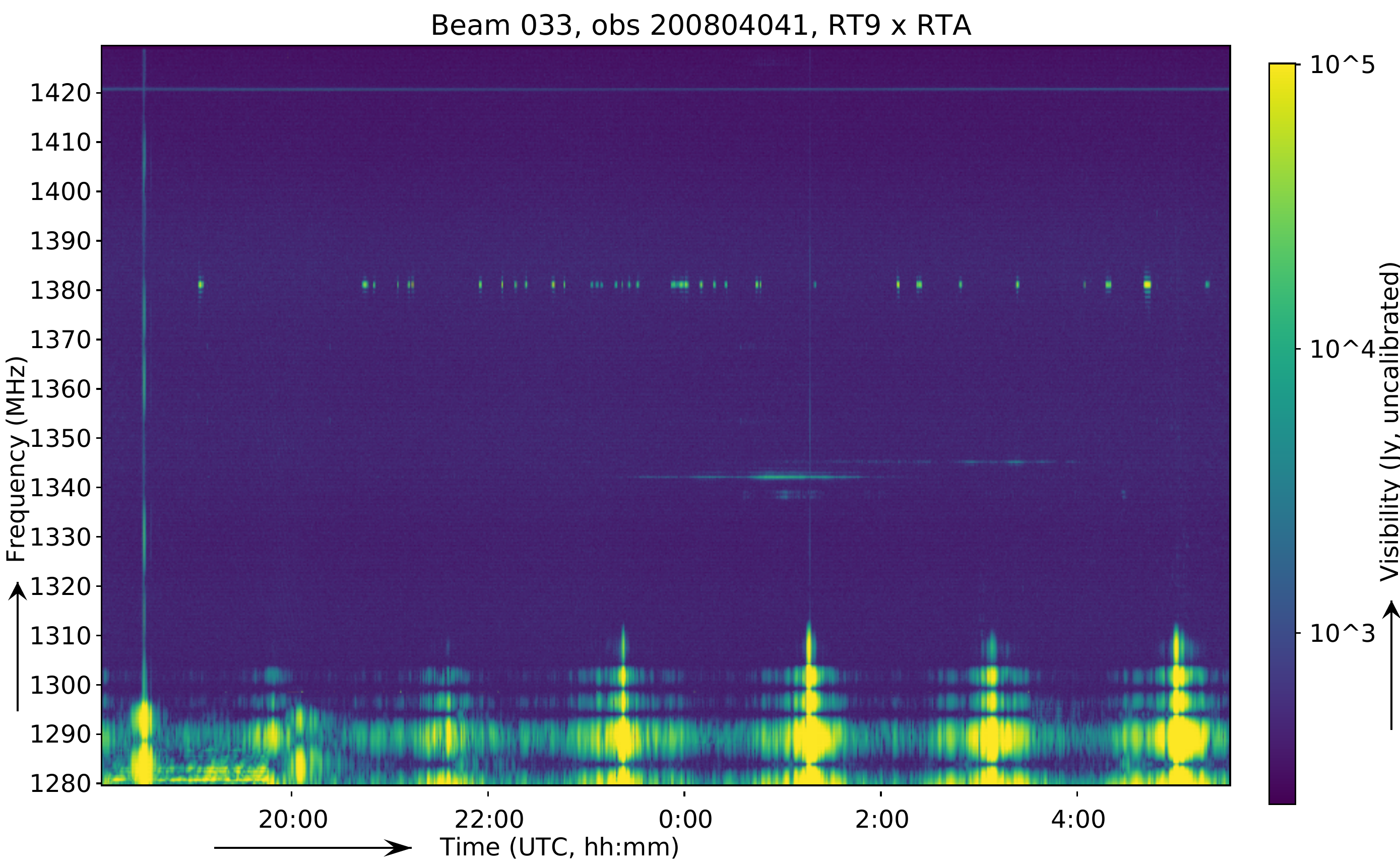}
		\end{subfigure}%
		\begin{subfigure}[c]{0.5\textwidth}
			\includegraphics[width=0.98\columnwidth]{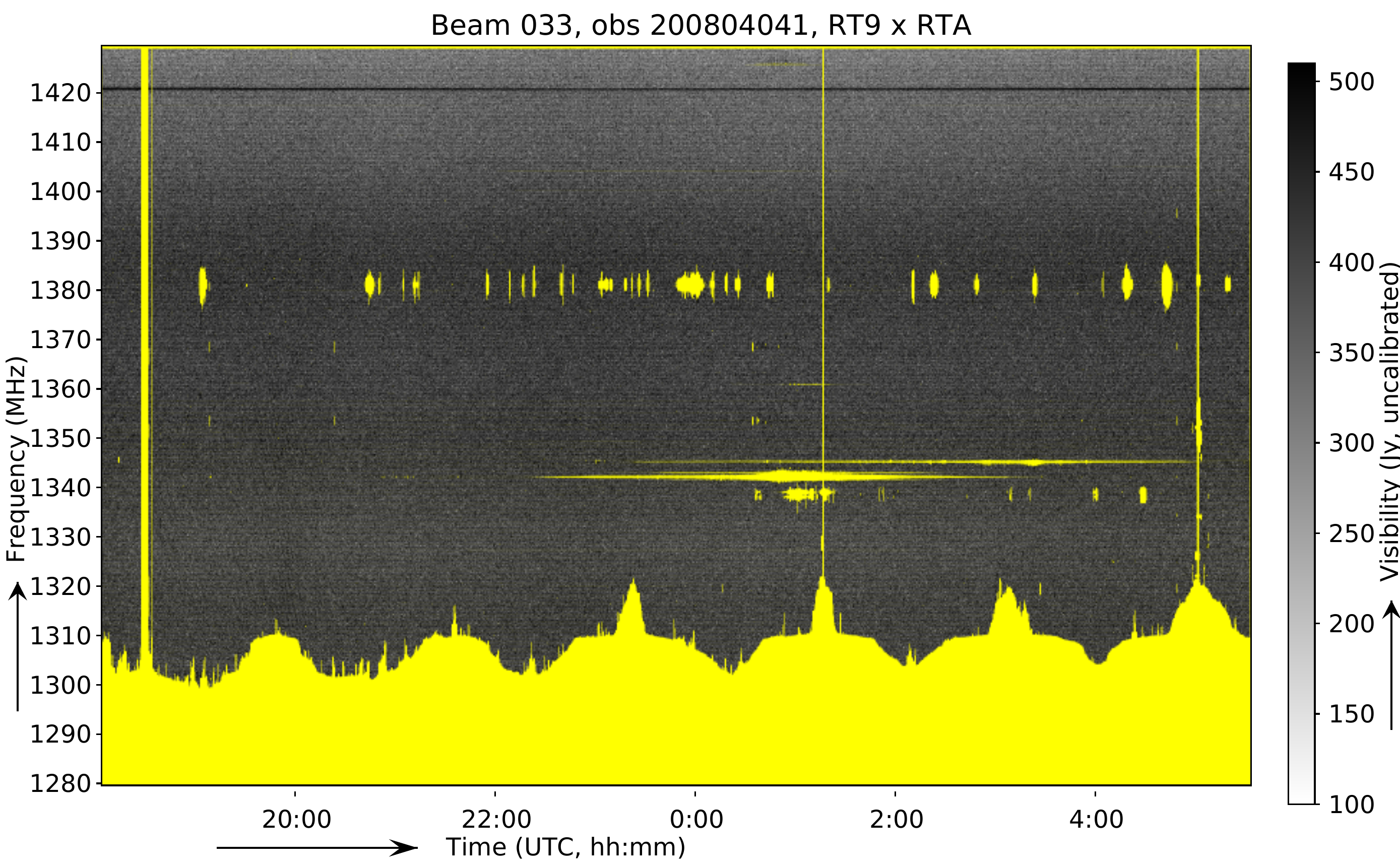}
		\end{subfigure}
		\caption{Time vs frequency plots of a typical Apertif 11.5\,hr integration. The scale is representing the amplitude of the visibilities. The left hand side shows the original data without any flagging. Automatic flags determined by AOFlagger are shown on the right hand side for the same data coloured in yellow. The HI-line at 1421\,MHz is successfully omitted during the flagging.}
		\label{image_RFI}
	\end{figure*}
	
	
	The designed AOFlagger strategy for Apertif makes use of a low-pass filter in combination with the Sumthreshold and Scale Invariant Rank operations to detect interference \citep{2010MNRAS.405..155O,2012A&A...539A..95O}. Previously set flags written to the MS and are used by AOFlagger for flagging the data further. An example of our strategy applied to Apertif data is shown in Fig. \ref{image_RFI}. Details and further results of the AOFlagger strategy will be discussed in Offringa et al. (in prep.). 
	
	\subsection{Crosscal}
	\label{subsect_crosscal}
	
	Cross-calibration is performed independently for each cali-brator-target-pair of observations. The cross-calibration step solves for the bandpass, gain, delay and polarisation leakage solutions of the flux calibrator. Because the flux calibrator is unpolarised, the cross-hand delay and polarisation angle solutions are derived from the polarised calibrator. All \emph{crosscal} calibration steps use the standard CASA routines in release version 4.7.0-1-el7. In case a polarisation calibrator has not been successfully observed or its dataset has not passed the \emph{preflag} module, polarisation leakage, polarisation angle and cross-hand delay solutions are not determined. This criterion is evaluated on a per-beam basis. In such a case, later stages of the pipeline omit polarisation imaging, but still perform self-calibration, line and continuum imaging to maximize the amount of the generated data products. Bandpass, polarisation leakage and polarisation angle solutions are derived on a per-channel basis to mitigate any effects within the observed bandwidth. For the unpolarised calibrators the flux density scale from \citet{2017ApJS..230....7P} is used while for the additionally needed information for the polarised calibrators, such as the polarisation angle, degree of polarisation and rotation measure, \citet{2013ApJS..206...16P} is used. We want to note, that for linear feeds CASA is not properly solving for polarisation angle corrections. This results in an exchange of the Stokes Q- and U-values for the two polarised calibrator sources used. This behaviour was found to be constant in time and over frequency, so that a correction can be performed after imaging the polarisation data products. Therefore, it is carried out during the polarisation mosaic step (see \ref{subsect_polarisationimaging}).
	
	Calibrator data are automatically checked for problematic dish-beam combinations. Problems here arise from individual receiver elements in the PAF, which are malfunctioning due to broken connectors, cables, or electronics. Beams are formed in a certain direction by weighting the signal of the different elements accordingly \citep{2021arXiv210914234V}. Beams where the weights for malfunctioning elements are high, result, if used for calibration and imaging, in higher noise levels in the images or even a complete divergence of the calibration solutions. These problematic beams are spotted most easily in the auto-correlation data. The amplitude and spectral behaviour of the Apertif auto-correlations is known well, so that differences from the expected values were used to develop a metric for an automatic detection algorithm.
	
\begin{figure*}
		\resizebox{\textwidth}{!}{\includegraphics{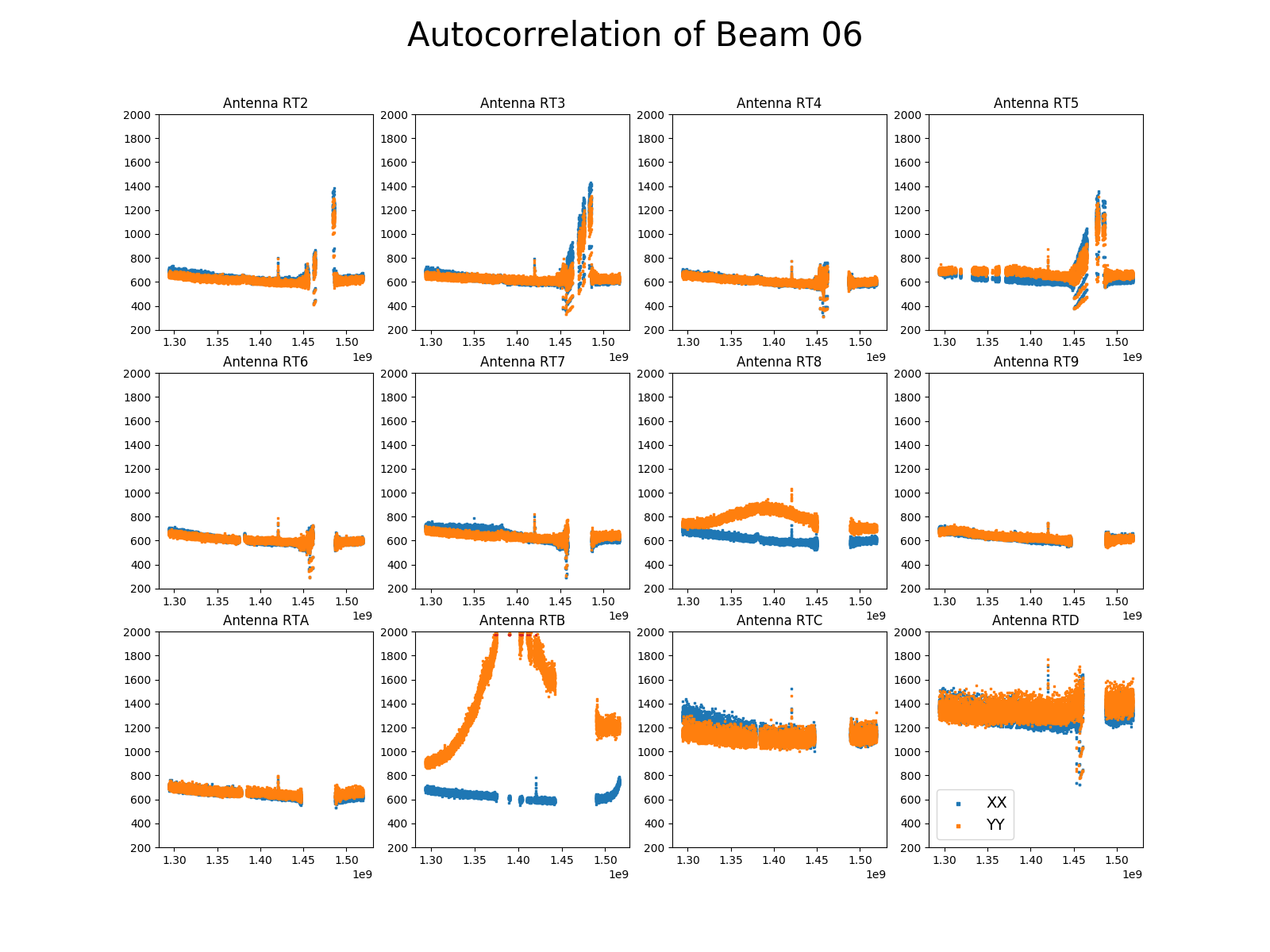}}
		\caption{Example plot for the quality assurance pipeline. Amplitudes of the autocorrelation visibilities are shown vs frequency for Beam 06 of one observation. Each subplot shows the XX (orange) and YY (blue) autocorrelations for one antenna. In this example the YY-autocorrelation values for antenna RTB show an atypical slope and curvature and more than 50\,\% of the values are larger than 1500\,K, so that all of its visibilities were flagged for this beam. RTC and RTD show constantly higher values, but both antennas are below 1500\,K, so that we were still able to use the data from these antennas.}
		\label{image_Autocorrelation_Tsys}
	\end{figure*}

	The currently implemented metric checks on a per beam and dish basis for the flux calibrator after a first cross-calibration, that not more than 50\% of the auto-correlation data show amplitudes of more than 1500\,K (see Fig. \ref{image_Autocorrelation_Tsys}). This value was derived from our experiences, where significant artefacts in the images become apparent. In addition, the bandpass phase solutions of the flux calibrator are investigated after calibration. A standard deviation of more than $15\deg$ is marking the calibration as failed. If one of the above mentioned criteria applies to a dish-beam combination, the specified data is marked and flagged automatically. The flags are then extended to the rest of the datasets of the observation, namely the target and polarisation calibrator data. This whole process is performed in an iterative way, where the whole cross-calibration is redone and results are checked by the metric described above. A maximum of four crosscal iterations are allowed after which the \emph{crosscal} module gives a final result. The pipeline is stopped for beams not passing this stage so that further processing on that particular beam ceases, but other beams passing the stage continue to be processed through the pipeline. Beams passing these checks are used to calibrate the target fields by applying their calibrator solutions to the target field data. Since the Apertif system does not possess a diode injected gain correcting system as the old WSRT MFFE backend did, the stability of the crosscal solutions was checked by applying different calibrator solutions several days apart from each other to target datasets. None of the final datasets showed systematic offsets or differences of target field source fluxes by more than 10\,\%. Any further processing of the calibrator datasets stops here and the following modules only focus on the target data.
	
	\subsection{Convert}
	\label{subsect_convert}
	
	Based on the familiarity of the developers with suitable software packages for the further reduction steps combined with the experience gained from the reduction of archival WSRT data using mostly MIRIAD, we chose to stay with this reduction package. We used the native ATCA MIRIAD version with slight modifications to be able to read and calibrate the Apertif data (see \ref{appendix_miriad}).
	
	Since MIRIAD is not able to access the Measurement Set\footnote{\url{https://casa.nrao.edu/Memos/229.html}} (MS)-format native to CASA, we need to convert the file format before doing any further reduction. Unfortunately a task for a direct conversion from MS to MIRIAD format is not available, so that we have to first convert to the UV-FITS standard \citep{Cotton1990} and from there to MIRIAD. For this purpose we use the CASA task \textsc{exportuvfits} followed by the MIRIAD routine \textsc{fits}.
	
	\subsection{Selfcal}
	\label{subsect_selfcal}
	
	Self-calibration is a standard procedure in radio interferometric data reduction to enhance the dynamic range of images. Small changes in the processing of the signals in the receiver electronics (e.g. temperature changes) and ionospheric and tropospheric variations of the Earth's atmosphere cause changes of the received amplitudes and phases over the time of the target observations. These usually slowly changing variations cannot be compensated by the bracketing calibrator observations and therefore need self-ca\-libration. This technique, developed more than 40 years ago and nicely summarised in \citet{1984ARA&A..22...97P}, involves the calibration of the data by using a model generated by imaging the data itself. 
	
	The task of the \emph{selfcal} module is to solve for the antenna and feed time-based variations of the target data within a self-regulating algorithm using the self-calibration technique. To guarantee the stability of the self-calibration process and the processing within a reasonable time frame, several preliminary steps are executed within the \emph{selfcal} module before the actual self-calibration starts.
	
	First, the target data is averaged down in frequency by a factor of 64 over the 64 channels of each subband resulting in a frequency resolution of 0.78\,MHz and therefore a bandwidth smearing of a $<1\%$-level. We also do not expect any strong amplitude or phase variations within this frequency span due to ionospheric variations or from the receiver hardware. It is important to note that this frequency  averaged dataset is only used for continuum and polarisation imaging in later stages of the pipeline and any HI-line imaging is performed on a dataset with the original resolution where the derived self-calibration gains are interpolated in frequency and applied.
	
	In order to mitigate any influence of strong HI-line emission or residual RFI on the self-calibration solutions we generate an image cube out of the averaged data. For each of the 64 images in the cube its standard deviation is measured. We then use an outlier detection algorithm to locate the channels affected by either above mentioned reasons and flag them in the averaged dataset. As above, these flags are only used for the continuum and polarisation imaging later in the pipeline and not for HI-line imaging.
	
	The performance of the self-calibration is often strongly dependent on the first image passed to the solver. In order to start the self-calibration with an image of an optimal initial quality, we use the information provided by radio continuum surveys at the same wavelength. For this purpose we first query the catalogue of the Faint Images of the Radio Sky at Twenty-Centimeters (FIRST) Survey \citep{1995ApJ...450..559B}. Since this survey does not cover the whole Apertif survey footprints information for fields outside of the FIRST footprint are collected from the NRAO Very Large Array Sky Survey (NVSS) \citep{1998AJ....115.1693C}. We observe with a fractional bandwidth of $\approx 20\%$, so that we need to account for the spectral index and primary beam variations over frequency. For acquiring a spectral index for the sources in our skymodel we query the Westerbork Northern Sky Survey (WENSS) catalogue \citep{1997A&AS..124..259R} and cross-match. Since WENSS has inferior resolution compared to the other two surveys, we account for multiple source matches by summing the fluxes of the individual components to derive the spectral index and assign the same value to all of them. We then account for the primary beam response of the Apertif system by using the primary beam model of the WSRT before the Apertif upgrade as an approximation \citep{2008A&A...479..903P}. The final skymodel is then generated by directly fourier transforming the catalogue source fluxes and positions into the (u,v)-domain with the MIRIAD task \textsc{uvmodel}. We then derive phase-only calibration solutions, which ensures that all our images are aligned to the same common reference frame given by the above mentioned surveys. In addition the resolution of this parametric skymodel is not limited by the pixel raster of the images, but rather the fitted position of the sources. The solution interval for this parametric calibration is usually on the order of several minutes, which is set in the configuration file. We performed several tests on this method by comparing sources detected in finally self-calibrated images of Apertif observations with archival values of the NVSS- and FIRST-catalogues and archival WSRT-observations. Differen\-ces were always on the order of a only few percent with no obvious systematic effects. A quantitative analysis of source position and flux accuracy for the Apertif surveys will be published in Denes et al. in prep. and Kutkin et al. in prep.
	
	The next step involves the actual self-calibration iterations. To improve the stability of the pipeline a multi-frequency image of circular polarisation (Stokes V) is generated at the beginning of each iteration. Since the circular polarised sky is essentially empty any sources in such an image would hint to severe calibration problems. Therefore, the image statistics can be analysed for following a normal distribution resembling Gaussian noise. This was implemented using the method developed by \citet{10.1093/biomet/58.2.341} and \citet{10.1093/biomet/60.3.613}, which combines information of the skewness and kurtosis. If these values exceed a certain number given in the configuration file, the self-calibration is aborted.
	
	After passing this check each iteration consists of inverting the (u,v)-data using the MIRIAD task \textsc{invert} followed by an automatic masking routine involving the source finder PyBDSF \citep{2015ascl.soft02007M} to limit the CLEAN algo\-rithm \citep{1974A&AS...15..417H} to islands of emission from astronomical sources omitting any image artefacts. Within PyBDSF we use an adaptive rms box without a pre-generated rms map. The rms value up to which PyBDSF is limiting the source detection is adaptive and is described below. All other parameters are set to their default values.
	
	We are aware that for linear feeds the Stokes Q emission can have a non-negligible effect on the calibration. Denes et al. in prep. showed that the contribution (leakage) for nearly all beams is only on the order of 1\,\% up to a primary beam response level of $\sim30$\,\%. Therefore, we perform all self-calibration and imaging is performed on the total intensity Stokes I data only. Image deconvolution is executed using the multi-frequen\-cy CLEAN-algorithm \citep{1994A&AS..108..585S} implemented into the MI\-RIAD task \textsc{mfclean}, which uses a first order polynomial to derive the spectral index of the sources within the imaged bandwidth. After cleaning, restored and residual images are created using the task \textsc{restor}. The CLEAN model generated during the cleaning process is then used to derive new calibration solutions.
	
	The CLEAN algorithm performs best for images which consist of point-sources and do not show calibration artefacts. In order to omit these artefacts and correctly generate masks for extended sources we use an iterative approach, where we set adaptive thresholds and constantly perform quality assurance incorporating different metrics as described in the following.
	
	Once one of the adaptive threshold criteria is reached the cleaning process is stopped. For each full cleaning iteration within a self-cali\-bration cycle three different thresholds for generating masks are calculated: the theoretical noise threshold $T_{tn}$, the noise threshold $T_n$ and the dynamic range threshold $T_{dr}$.
	
	The theoretical noise $T$ in Jy/beam is determined by calculating the standard deviation from images generated in circular polarisation (Stokes V). Astronomical circular polarised sources are very rare on the sky and if present only at very low flux levels. Residual RFI on the other hand is often circular polarised and raises the noise levels of these images. The noise statistics of these images therefore represent the actual quality of the data and the theoretically reachable noise of the final images well. $T_{tn}$ is given in units of Jy/beam and defined as
	
	\begin{equation}
	T_{tn} = T \cdot n_\sigma
	\end{equation}
	
	where $n_\sigma$ is the confidence interval for regarding islands of emission as real. This is usually set to $n_\sigma=5$.
	
	In order to guarantee a smooth convergence of the self-calibration skymodel the two additional thresholds $T_{dr}$ and $T_n$ set limits for the maximum dynamic range achievable in an image without reconstruction and the adaptation to image artefacts, respectively. The dynamic range threshold within a cycle is defined by the number of the current major cycle $m$, the initial dynamic range $DR_{i}$ and a factor defining how fast the threshold should increase $DR_0$ such as
	
	\begin{equation}
	T_{dr} = \frac{I_{max}}{DR_{i} \cdot DR_0^m} 
	\end{equation}
	
	where $I_{max}$ is the maximum pixel value in the residual image of the previous cycle. The parameter $DR_{i}$ is dependent on the level of the first major sidelobe in the dirty beam. The ratio between the maximum and this value gives the maximum dynamic range by which an image can be cleaned before another cycle of image reconstruction needs to be performed.
	
	The adaption of the noise threshold $T_n$ for stopping each individual run of the CLEAN algorithm is given by
	
	\begin{equation}
	T_n = \frac{I_{max}}{((c_0 + n \cdot c_0) (m + 1))}
	\end{equation}
	
	where $m$ and $n$ are the number of major and minor cycles, respectively, and $c_0$ handles how aggressive the cycles are performed. For each individual run of \textsc{mfclean} all three thresholds are calculated and the maximum is set as a limit for the generation of masks in PyBDSF for the current major cycle. The maximum number of major cycles is set in the configuration file. This is useful for cases where the theoretical noise threshold is approached very slowly or never reached due to remaining image artefacts. Then cleaning is performed within this mask down to a level of the derived masking level divided by the parameter $c_1$, which is usually set to $c_1=5$.
	
	The length of the solution interval $s$ for each self-calibra\-tion cycle is reduced for an increasing number of major cycles $m$. This is possible due to the improved and more complex CLEAN component model used and the better starting values for fitting solutions derived from earlier selfcal iterations. In addition a higher number of data points due to more baselines, feeds and solution intervals over frequency guarantee a better convergence of the self-calibration. Therefore, the solution interval is determined using the following equation:
	
	\begin{equation}
	s = \frac{t / n_s}{m}
	\end{equation}
	
	where $t$ is the total observation time and
	
	\begin{equation}
	n_s = \sum_{l} \frac{I(l)}{SNR \cdot T \sqrt{n_B n_{nf} n_f}}
	\end{equation} 
	
	where $l$ is the number of clean components, $I$ the flux of each individual clean component, $SNR$ the needed signal-to-noise ratio, $T$ the theoretical noise, $n_B$ the number of baselines, $n_{nf}$ the number of solution intervals over frequency and $n_f$ the number of polarisations to solve for. Due to the larger number of degrees of freedom for amplitude and phase calibration in contrast to phase-only calibration the $SNR$ is set to 3 for phase-only calibration and to 10 for combined amplitude and phase calibration. The above formalism ensures that solution intervals decrease during the self-calibration process consecutively while still containing enough signal-to-noise for a proper calibration.
	
	The \emph{selfcal} module first performs up to a given maximum number of iterations (typically five) of phase-only self-calibra\-tion. By summing the flux of the clean components the module calculates the total flux in the dataset and can estimate the available $SNR$. This allows to perform an additional self-calibration step for a combined amplitude and phase solution for datasets with high SNR improving their final image quality. The image quality of observations with low SNR would suffer from this step and are therefore excluded. If at any point during the process the theoretical noise limit is reached, \emph{selfcal} performs only one last iteration of self-calibration.
	
	\begin{figure*}
		\resizebox{\hsize}{!}{\includegraphics{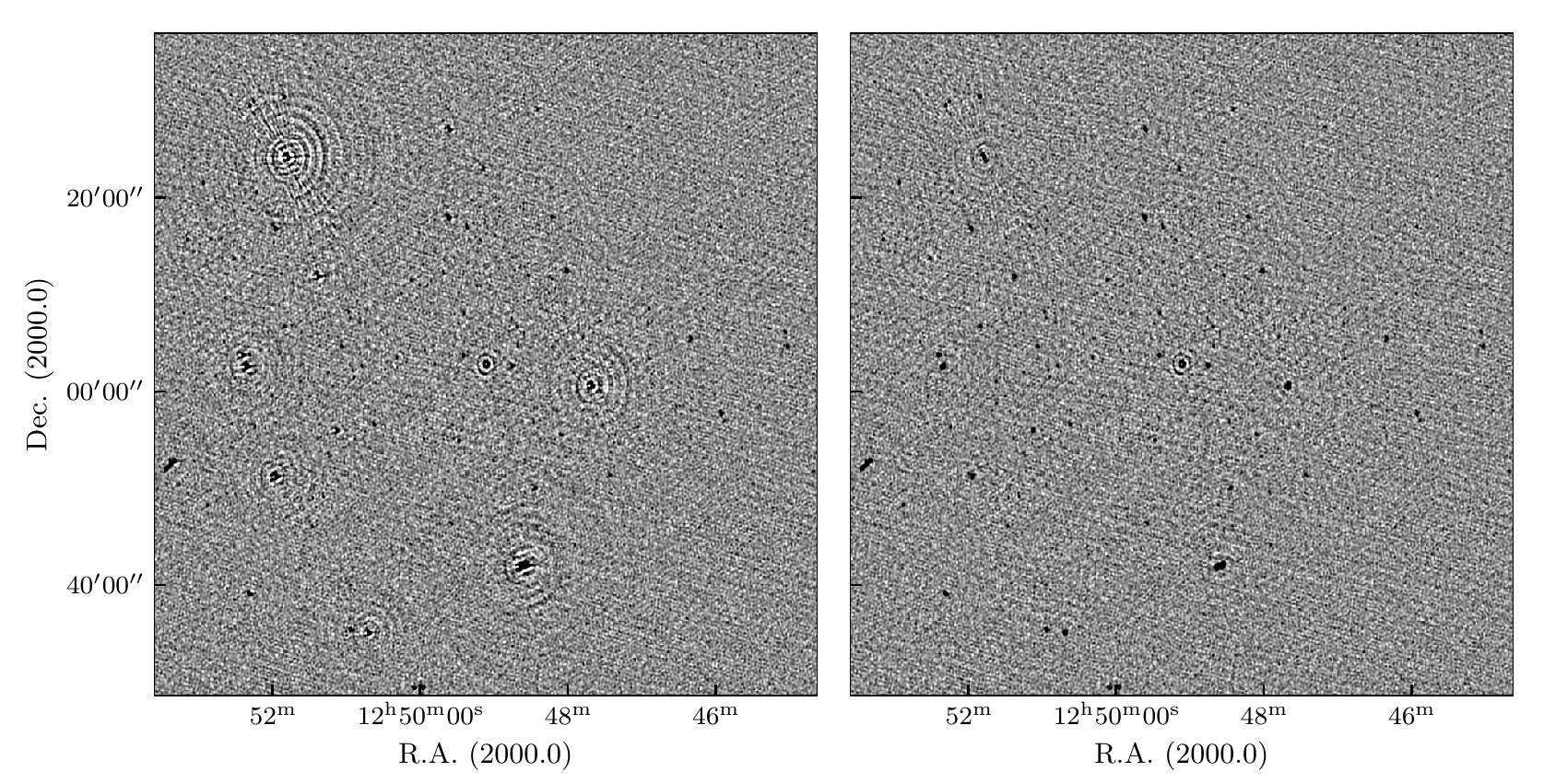}}
		\caption{Example of a continuum image before phase self-calibration (left) and after (right). Artefacts around stronger sources are significantly reduced and all sources appear more focused.}
		\label{image_selfcal}
	\end{figure*}
	
	During and after each imaging and cleaning cycle the dirty image, the cleaning mask, the clean component model image and the restored images are checked for any obvious problems resulting from a divergence of the calibration routines using the following criteria and metrics. The maximum value in dirty images of total intensity should always exceed the minimum. In addition images are inspected for Not a Number (NaN)-values. Both conditions are checked when starting a cleaning where new solutions were derived and applied to the data and a new dirty image is generated. Clean component images are checked for containing zero CLEAN components. In addition, the clean component images are checked for clean components with unrealistically strong negative or positive fluxes. The restored image is again checked for containing no NaN-values and for unrealistically strong positive or negative values on the order of $10^4$Jy. These NaN-values are generated in the rare cases where PyBDSF, during the masking procedure, does not find islands of emission and CLEAN of MFCLEAN are executed with an empty mask. This can often happen in cases where the data is suffering from system failures or diverging calibration solutions. The final residual image should mostly consist of noise and is therefore checked for gaussianity using the aforementioned algorithms .
	
	A combined amplitude and phase calibration (A\&P) is not as stable as a phase-only calibration due to the increased degrees of freedom, so that this step can easily worsen the image quality due to diverging calibration solutions. Therefore, in addition to the quality assurance process described above, we added an additional metric to check the quality of the A\&P calibration compared to the phase-only one. After the A\&P calibration we generate another dirty image and compare the image statistics independently, namely its maximum, minimum and standard deviation with the dirty image of the last phase-only self-calibration cycle. If the ratio of one of those values exceeds a limit given in the config-file, which is usually based on a worsening of the image quality after a combined A\&P-calibration step, the A\&P calibration is assessed as failed and any following module will use the last successful phase-only self-calibration solutions. Calibration solutions are applied in the subsequent modules before any further imaging is performed. An example of a final total power self-calibrated image in comparison to one without self-calibration is shown in Fig. \ref{image_selfcal}.
	
	\subsection{Continuum}
	\label{subsect_continuum}
	
	The \emph{continuum} module performs two different tasks to generate final deep continuum Stokes I images from the self-calibra\-ted (u,v)-data. First it generates a deep multi-frequency image using the \textsc{mfclean} task in MIRIAD and secondly several individual images spanning narrower frequency ranges of usually $\approx$20\,MHz over the full bandwidth using the task \textsc{clean} are produced. The purpose of this is to generate an as-deep-as-possible total intensity image with the maximum possible resolution (given by the highest frequency) using the multi-fre\-quency imaging approach and in addition derive reliable spectral indices and curvatures for as many sources as possible using the individual frequency images. In fact multi-frequency cleaning generates these images already, but their values are only reliable in cases of high signal-to-noise ratios. For both imaging steps we use a natural robust weighting \citep{1981MNRAS.196.1067C} to acquire the maximum possible resolution. Images generated with the default parameters have a size of 3073x3073 pixels with a sampling of 4'' per pixel, which allows the imaging and cleaning of any sources up to the first sidelobe level of the primary beam response in order to minimise artefacts. 
	
	Cleaning and masking iterations are in both cases continued until the theoretical noise limit has been reached. Masking and validation of all continuum images is performed in the same way as described in Sect. \ref{subsect_selfcal}.
	
	\subsection{Line}
	\label{subsect_line}
	
	The \emph{line} module first applies the derived self-calibration solutions to the non-averaged data. This is performed using the MIRIAD task \textsc{gpcopy}. It automatically takes care of the different frequency resolution of the two datasets by interpolation.
	
	The HI-line imaging is the most computing intensive task in the Apertif data reduction. The following strategy for handling the line data and the image cubes has been found to be most time efficient: imaging is performed by generating eight individual cubes over the 300\,MHz of bandwidth with a small amount of overlap in frequency. The overlap is necessary to avoid splitting the detected line emission of individual objects between two adjacent image cubes. In order to improve sensitivity and save processing time and disc space, data between 1130\,MHz and 1416\,MHz are averaged in frequency by binning three channels together resulting in a frequency resolution of 36.6\,kHz corresponding to a velocity resolution of 7\,km/s, which is usually sufficient for any extragalactic objects at these frequencies. The data at the highest frequencies which features the Galactic neutral hydrogen and small galaxies in the nearby Universe retain the full spectral resolution of 12.2\,kHz or 2.3\,km/s. All line data is in topocentric velocities, which is the native reference frame of the data. Conversions to other reference frames are performed during further analysis steps.
	
	We want to note that the current flagging and imaging strategy is not optimised for calibrating and imaging the Galactic range of the data. This would need a joint deconvolution of several beams at the same time plus a combination with single dish data, for example from the HI4PI-survey \citep{2016A&A...594A.116H}, to mitigate the effect of missing short spacings.
	
	In order to generate image cubes containing only HI-line emission the continuum has to be subtracted. Several different approaches are possible here: the fitting of baselines to the amplitude of the data followed by subtraction, the subtraction of constant fluxes over frequency in the image domain and the direct subtraction of a continuum clean component model from the (u,v)-data. The best performance in terms of time consumption and fidelity was achieved with the last method, so that we decided to use this in \emph{Apercal}. For this subtraction the final clean component model of the \emph{continuum} module is used.
	
	The final line images are produced using a parallelisation approach where each individual frequency channel is imaged as an individual imaging task. For this most computing intensive part of the pipeline, the Python Pymp\footnote{\url{https://github.com/classner/pymp}} utilisation of the OpenMP\footnote{\url{https://www.openmp.org/}} framework was used. All resulting images are then combined into single line image cubes.

	\subsection{Polarisation}
	\label{subsect_polarisation}
	
	Polarisation imaging is performed in Stokes Q, U and V. Q and U fluxes from astronomical sources exhibit a sinusoidal dependence of the square of the observed wavelength, which is caused by Faraday rotation. In addition, Stokes Q, U and V fluxes can have negative values in contrast to Stokes I, which needs to be positive in all cases. These effects would lead to bandwidth depolarisation in case of multi-frequency imaging of Stokes Q and U over our full 300\,MHz bandwidth \citep{1966MNRAS.133...67B}. Therefore, we image Stokes Q and U as cubes, using the method described in Sect. \ref{subsect_line}, where one image is generated for a bandwidth of 6.25\,MHz, which results into 48 images (subbands) for each Stokes parameter. This mitigates the effect of bandwidth depolarisation for most astronomical sources. In addition, this method allows the usage of the rotation measure synthesis technique \citep{2005A&A...441.1217B} in post-processing, so that the linear polarisation properties of the detected sources can be analysed in more detail than with the standard methods, which suffers from bandwidth depolarisation effects.
	
	Stokes V is representing the circular polarisation, which does not show any sinusoidal behaviour. Due to this fact and since the circular polarised sky is very faint, we perform a multi-frequency synthesis for imaging Stokes V. This also allows to maximise the sensitivity of the produced images to detect possible circular polarised sources.
	
	For all polarisation images cleaning is performed using the final mask generated by the multi-frequency imaging part of the \emph{continuum} module. Since polarisation images usually only show very faint emission, the clean threshold is set by the standard deviation of the pixels in the image. This accounts, especially for the Stokes Q and U images, for the variations of the noise over the imaged bandwidth. Since we are only imaging a limited bandwidth at a time for Stokes Q and U we use the faster MIRIAD task \textsc{clean} while for the imaging of the full bandwidth for Stokes V the task \textsc{mfclean} is used. We want to note that due to the separate cleaning of Stokes Q- and U-images biases regarding the polarisation angles could be introduced into the final maps as has been shown by \citet{2016MNRAS.462.3483P}.
	
	\subsection{Mosaic}
	\label{subsect_mosaic}
	
	Once all data of an observation have been processed through the \emph{continuum}, \emph{line}, and \emph{polarisation} modules, a \emph{mosaic} module can be executed to generate a combined image of all beams of one observation. This step is currently not performed as part of the automatic pipeline.
		
	Images are regridded to a common grid centred on the central beam of the observation and then corrected for the Apertif compound beam response. The compound beam response has been characterised using drift scans of a strong astronomical source over the whole field of view of the Apertif Phased Array Feed (for details see Denes et al. in prep.). The different beams of one observation have slightly varying synthesised beam sizes due to different flagging of the data and slightly different declination positions, so that all input images are convolved to the a given common beam.
	
	The combination of the input images then follows an inverse square weighting based on the compound beam response and the background noise of the individual images. The background noise is estimated using the MIRIAD task \textsc{sigest}, which minimises the contribution of sources for the determination of the noise level.
	
	Since all data of one observation has been taken using the same electronics, a correlation between the noise of different beams exists, which raises the noise level of the final mosaic. An option for including this correlation matrix during mosacking is implemented and will be used once the coefficients have been measured. First tests showed a minor change of the noise levels between correlated and uncorrelated data of adjacent beams of $\sim 2\,\%$.
	
	The \emph{mosaic} module is currently only producing continuum mosaics using the central frequency of the observational setup for correcting the primary beam response. Additional features in the future include the implementation of the frequency and long-term time dependence of the beam pattern. The current implementation takes approximately an hour for generating a continuum mosaic. Future improvements will include enhancements to the speed of the module, which will then allow us to generate polarisation and line mosaics within an acceptable amount of time.
	
	\subsection{Transfer}
	\label{subsect_transfer}
	
	This module converts the self-calibrated MIRIAD data to UVFITS format. Similar to \emph{line} it applies the phase and, if available, amplitude self-calibration solutions to the non-avera\-ged cross-calibrated data automatically before conversion. The data is then ready for archive ingest.
	
	\section{Additional mosaicking software}
	\label{sect_mosaicking}
	
	To allow users a convenient workflow with the provided Apertif data products, especially continuum images and polarisation and line cubes, we developed the \emph{Apercal} independent mosaicking package \emph{amosaic\footnote{\url{https://github.com/apertif/amosaic}}}. This package is purely based on up-to-date Python3 routines and allows command line or Jupyter notebook execution. These specifications ensure that the astronomical community is able to work with the provided data products of the current and future Apertif data releases without the need for installing or maintaining any radio astronomical calibration packages.
	
	The current version of the software is able to mosaic Stokes I continuum images and Stokes Q and U cubes of single observations. In the following we want to describe the capabilities of the software and give an outlook on future improvements.
	
	\subsection{Continuum mosaicking}
	\label{subsect_continuumimaging}
	
	Continuum images for the mosaicking operation are accepted or rejected based on their noise, synthesised beam parameters and different quality metrics (see Adams et al. in prep for a detailed explanation). Accepted images are then primary beam corrected using externally supplied primary beam models. A description of the two different methods to measure the Apertif primary beam are given in Denes et al. in prep. and Kutkin et al. in prep.. Primary beam response images are adjusted to the frequency and size of the target image and then multiplied.
	
	We determine the smallest common beam of all accepted images using the Python package \emph{Radio Beam\footnote{\url{https://radio-beam.readthedocs.io/en/latest/}}} with small adjustments to the released code for better handling of small convolution kernels. The Python package \emph{reproject\footnote{\url{https://reproject.readthedocs.io/en/stable/}}} is used to convolve all images to the calculated common beam.
	
	For the combination of the images we use the \emph{reproject\_interp} algorithm inside the \emph{reproject} package. Images are inversely square weighted according to their noise characteristics. The final mosaic is then clipped to a given level of the primary beam response and written to disc.
	
	\subsection{Polarisation mosaicking}
	   \label{subsect_polarisationimaging}
	
    Polarisation mosaicking is performed as described in Sect. \ref{subsect_continuumimaging} with adjustments needed for handling image cubes and polarisation specific analysis steps. Our calibration strategy introduces an offset in the polarisation angle calibration (see Sect. \ref{subsect_crosscal}). To correct for this effect the final Stokes Q and U images are exchanged with each other. Then the image cubes are prepared for further processing in view of the RM-Synthesis technique \citep{2005A&A...441.1217B}. Since this technique is mathematically a Fourier Transform which best works on regularly gridded data, all images in the cubes need to be on the same frequency grid. In addition, a variation of the input synthesised beam along the frequency axis would cause spatial variations in the resulting Faraday cubes leading to additional unpredictable biases for the scientific analysis. We therefore generated a routine to illustrate the final accepted beams and channels for given constrains. Limits for the mosaicing rotuines can be set on noise in individual channel images, their synthesised major and minor beam sizes and the maximum images for a certain beam these constrains are not met for. This routine can easily be used to optimise the input parameters for polarisation mosaics to maximise the output with regard to sensitivity and resolution. The individual image mosaics are generated independently in Stokes Q and U using the same procedure as described in Sect. \ref{subsect_continuumimaging}. As a final step all image mosaics are combined into Stokes Q- and U-image cubes.
	
	\subsection{Future improvements}
	
	Additional features for the mosaicking software include the combination of different Apertif pointings. The mosaicking for line cubes is currently in development. Due to the large data volumes, which need to be processed, this also includes the parallelisation of the mosaicking software of different channels in the line image cubes at the same time. In addition, as described in Sect. \ref{subsect_mosaic}, noise between beams of the same observation is correlated. The calculation of these correlation coefficients and their application to the images is another future improvement.
	
	\section{Final data products and archive ingest}
	\label{sect_finaldata}
	
	Upon the successful completion of an \emph{Apercal} run, the pipe\-line reports back to \emph{Autocal} about which beams (if any) have failed on particular steps. All final data products need to be uploaded to ALTA on a beam-by-beam basis, and connected back to the original raw data via the target IDs. To guide this process, we built \emph{Apergest}, which manages the ingest of processed data products. \emph{Autocal} sends a trigger call to \emph{Apergest} for a given target ID, which results in \emph{Apergest} first referencing a list of expected data products which it then searches for on the cluster. If any data products are missing due to failures in particular steps of the pipeline, it simply continues the upload with what is available.
	
	Once \emph{Apergest} has identified all relevant files, it produces a file in JSON format with metadata information about each file which can be interpreted and translated by ALTA. We also include additional metadata, such as the \emph{Apercal} pipeline version or \emph{Apergest} version. The resultant JSON file is fed to an ALTA instance (which is run inside a Singularity container \citep{PLoSOne_singularity}) in order to prepare the ingest. Once successfully prepared, \emph{Apergest} starts the ingest which results in the actual upload of each data product. Finally, \emph{Apergest} then deletes all processed data products which have been successfully uploaded.  
	
	\section{Data quality assessment}
	\label{sect_dataquality}
	
	We built the Python package \emph{DataQA}\footnote{\url{https://github.com/apertif/dataqa}} for quality assessment (QA) after the data were processed with \emph{Apercal}. The aim of this approach is not to identify single problematic data products such as single images in a line cube or small outliers during calibration, but rather give an overall assessment of a complete observation and its reduced \emph{Apercal} products. Often a failed run of \emph{Apercal} or low quality imaging products can be traced back to a small amount of bad data and observations do not have to be repeated. The quality assessment of individual beams and the release of those for public usage will be described in Adams et al. in prep.
	
	\emph{DataQA} was developed specifically for this purpose. It produces diagnostics for each step of \emph{Apercal} to support us in assessing the quality of the processing and the observation. For easier access to all QA products, \emph{DataQA} produces an html-based overview which can be viewed in the web browser. In the following, the QA products for each \emph{DataQA} step are briefly described.\\

\begin{figure*}
		\resizebox{\hsize}{!}{\includegraphics{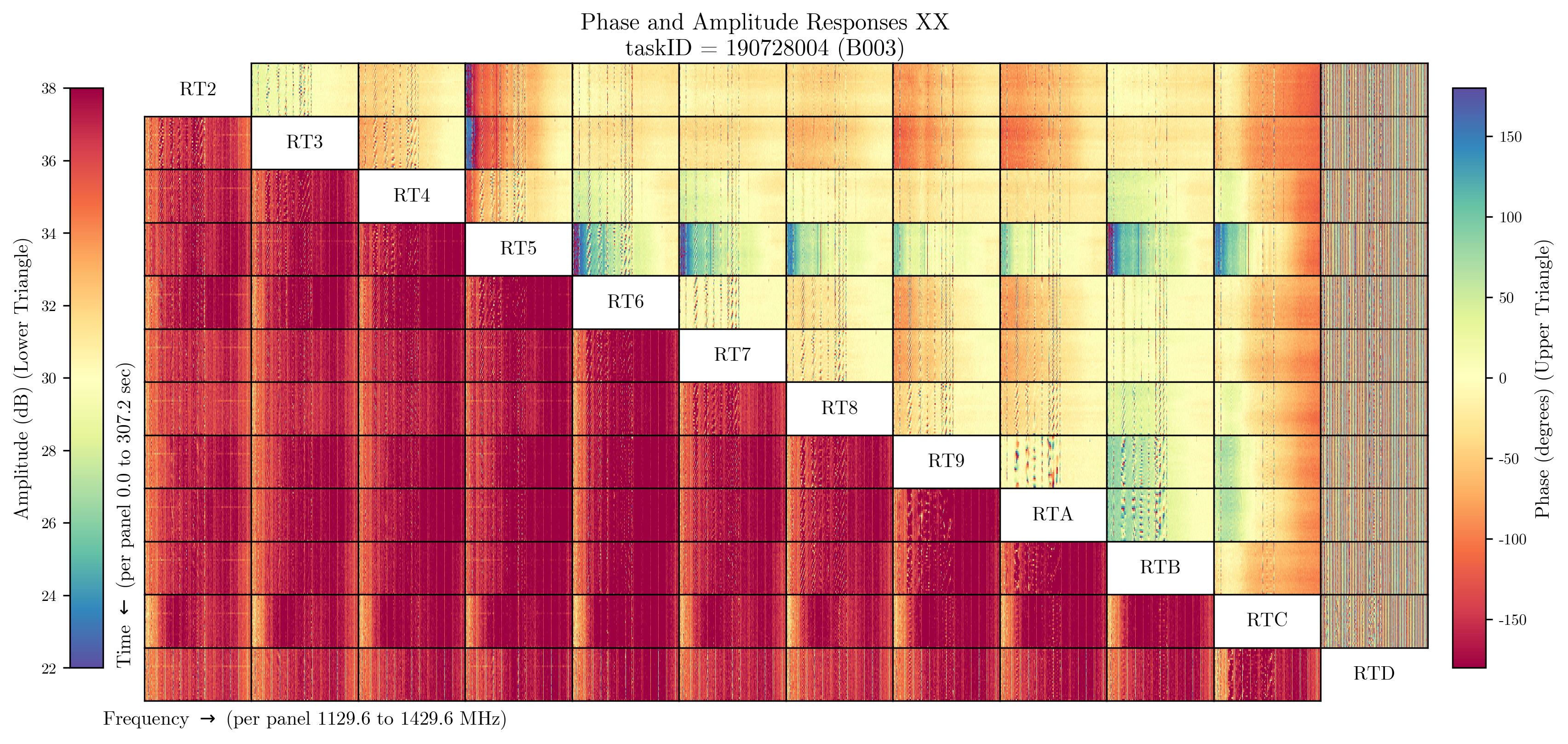}}
		\caption{Example plot of the raw uncalibrated data for the data quality assurance. Phase (top right triangle) and amplitude (bottom left triangle) are visualised as a colourscale. Each plot represents the visibilities of one baseline on a frequency vs. time basis. The difference in terms of RFI environment is clearly visible for the lower part of the observed freqeuncy band compared to the higher one. In this example antenna RTD was malfunctioning, which can easily be identified by the discontinuous phase behaviour.}
		\label{image_ALTAdiagnostics}
	\end{figure*}

	\noindent\emph{Prepare:} \emph{DataQA} retrieves diagnostic plots from ALTA that are based on the uncalibrated raw data. They are created for the target field and both calibrators for each beam of an Apertif imaging survey observation (see Fig. \ref{image_ALTAdiagnostics} for an example plot of a calibrator observation). These plots provide information on the impact of RFI, the quality of the tuning of the amplifiers in the system, the pointing accuracy as well as other potential issues with the system itself such as failing electronics during the observation.\\

	\noindent\emph{Preflag:} \emph{Apercal} produces plots of the RFI flagging performed by \emph{AOFlagger} for three baselines that represent a short, an intermediate and a long baseline of the WSRT (see Fig. \ref{image_RFI} for an example). The plots are produced for each beam of the calibrators and the target. \emph{DataQA} collects these plots and combines them for better inspection of the RFI flagging done across the different beams.\\
	
	\noindent\emph{Crosscal:} \emph{DataQA} generates plots for all derived cross calibration solutions in the \emph{crosscal} module (see Fig. \ref{image_Autocorrelation_Tsys} for an example). This allows an easy identification of malfunctioning frequency bands or dishes and residual RFI influencing the calibrator solutions.\\
	
	\noindent\emph{Selfcal:} For each self-calibration iteration and each beam the model, residual and restored images are shown in this part of the module. Due to the large amount of images, these are mostly used for closer inspection of certain beams. The most important usage of the images are the final model images. Due to image artefacts and the non-linear nature of the CLEAN-algorithm, sidelobes of sources are sometimes included in the cleaning masks. The CLEAN-components within these masks will then be considered as real sources for the following self-calibration iterations. These ring-like structures are easy to identify by eye. In addition to the individual images, the final phase and amplitude self-calibration solutions are plotted for each beam, time and frequency solution interval.\\
	
	\noindent\emph{Continuum:} The final clean component model images and restored and residual images are shown for each beam. The final quality of the images can easily be inspected by eye to spot any obvious artefacts in the images. For a quantitative quality assurance we use an \emph{Apercal}-specific version of the ASKAP continuum validation tool\footnote{\url{https://github.com/Jordatious/ASKAP-continuum-validation}}. This tool automatically identifies the stronger continuum sources in the final images and compares their position and fluxes to the NVSS database. This therefore serves as a check of the overall flux scale of the final images as well as their astrometry.\\
	
	\noindent\emph{Line:} A statistical approach is used for the quality assurance of the HI-line cubes due to the large amount of images, which would otherwise need to be inspected. The standard deviation for each individual image in the cube is determined and plotted vs. the frequency. This allows for an easy identification of individual low quality images as well as broad bands of missing data or images affected by broad band RFI.\\
	
	\noindent\emph{QA overview and quality report:} Currently for each observing week an observational support astronomer (OSA) is assigned the task of inspecting the data quality assessment plots. Each individual \emph{DataQA} step is then ranked on a scale between 0 and 6 based on the number of successfully reduced beams. Additional constraints encompass the loss of the long baselines, if dishes RTC and RTD failed, which enlarges our synthesised beam size by a factor of two. Individual comments can be inserted by the OSA. The complete report is saved to disk, so that all numbers can be accessed at a later point in time. 
	
	\section{Summary and outlook}
	\label{summary}
	
	Apercal is the dedicated data reduction pipeline for the imaging surveys of the SKA pathfinder telescope Apertif. We managed to create a data reduction framework using an object oriented Python-based approach. Well-established astronomical software packages are used to minimise the workload for the developers and optimise the timing and quality of the results. In addition, Apercal is partly parallelised using Python software packages, which make it easy to adjust to any multi-node hardware system. Currently Apercal is taking 24\,hrs for a 11.5\,hr target observation to reduce. Since only two thirds of the telescope time is dedicated to imaging observations and calibrator observations, beam weight measurements as well as maintenance produce an additional overhead of $\sim$25\,\%, it is possible for our setup to keep up with the incoming data amount.
	
	The modularity and independence of individual modules and routines allowed the usage and testing of the pipeline during commissioning of the Apertif system already. The whole pipeline and its prerequisites are publicly available, so that users within the astronomical community are able to improve and modify it to their needs.
	
	In order to ensure the quality of the generated images and the stability of the pipeline two stages of quality assurance are used. The first one being fully automatic and executed during the reduction of the data. Each step in the reduction process including statistics and prerequisites for the following steps is saved in a stats-file. This allows the user to directly see at which stage the data reduction was successful or failed. The second step of quality assurance is executed after each individual pipeline run. Here plots are generated and solutions and images are visualised in a compressed format, so that failing observations or bad quality images are noticed by an observer within an optimal timeframe. Currently 44\,\% of the observed data processed by Apercal are on a science-ready quality level.
	
	Future improvements to Apercal include an improved masking algorithm, an automatic direction dependent reduction step in full polarisation and direction dependent leakage corrections. First tests showed that these improvements increase the percentage of science ready data products to over 80\,\%. 
	
	\section{Acknowledgements}
	
	JMvdH and JV acknowledge support from the European Research Council under the European Union`s Seventh Framework Programme (FP /2007-2013), ERC Grant Agreement nr. 291531 (HIStoryNU). EAKA is supported by the WISE research programme, which is financed by the Netherlands Organization for Scientific Research (NWO). The research leading to these results has received funding from the European Research Council under the European Union's Seventh Framework Programme (FP/2007-2013) / ERC Advanced Grant RA\-DIOLIFE-320745. JvL, LDC, LCO, YM and RS acknowledge funding from the European Research Council under the European Union’s Seventh Framework Programme (FP/2007-2013)/ERC Grant Agreement n. 617199 (‘ALERT’). JVL and LCO were further supported through Vici research programme `ARGO' with project number 639.043.815, financed by NWO. DV acknowledges support from the Netherlands eScience Center (NLeSC) under grant ASDI.15.406. The Westerbork Synthesis Radio Telescope is operated by ASTRON (Netherlands Foundation for Research in Astronomy) with support from the Netherlands Foundation for Scientific Research (NWO). This research made use of Astropy,\footnote{http://www.astropy.org} a community-developed core Python package for Astronomy \citep{2013A&A...558A..33A, 2018AJ....156..123A}. This research made use of the Python Kapteyn Package \citep{KapteynPackage}.
	
	\begin{appendix}
	
	\onecolumn
	\section{Configuration file}
	\label{appendix_configfile}
	An example for a pipeline configuration file is shown below. Individual parameters are sorted in the usual execution order of the \emph{Apercal} modules. Parameter names begin with the name of the module where the parameter is introduced and used for the first time. Parameters can have any format, which is supported by the Python ConfigParser package.\\
	
	\vdots
	\begin{verbatim}
[CROSSCAL]
crosscal_refant = 'RT2'
crosscal_initial_phase = True
crosscal_global_delay = True
crosscal_bandpass = True 
crosscal_gains = True
crosscal_crosshand_delay = True 
crosscal_leakage = True
crosscal_polarisation_angle = True
crosscal_transfer_to_cal = True
crosscal_transfer_to_target = True
crosscal_autocorrelation_data_fraction_limit = 0.5
        
[CONVERT]
convert_fluxcal = True
convert_polcal = True
convert_target = True 
convert_removeuvfits = True
convert_removems = True
convert_averagems = True
	\end{verbatim}
	\vdots
	
	\section{Parameter file}
	\label{appendix_paramfile}
	Parameters saved within the parameter file for the \emph{preflag} module for a single beam (Beam 07) are shown below as an example. Parameters are saved as bibliographies and saved to disk as pickled numpy objects. The nomenclature for each parameter starts with the module name where it was generated, followed by the beam number and the actual parameter name. Parameter values can have all python supported formats and be of any desirable dimension. Helper functions to access or modify the parameter file can be found within the \textsc{param.py} subroutine in \emph{Apercal}.\\
	
	\vdots
	\begin{verbatim}
preflag_B07_targetbeams_manualflag_auto False
preflag_B07_fluxcal_manualflag_time ['']
preflag_B07_fluxcal_manualflag_baseline ['']
preflag_B07_fluxcal_manualflag_antenna ['']
preflag_B07_polcal_manualflag_from_file False
preflag_B07_targetbeams_edges True
preflag_B07_fluxcal_ghosts False
preflag_B07_fluxcal_edges False
preflag_B07_fluxcal_manualflag_auto False
preflag_B07_targetbeams_manualflag_clipzeros False
preflag_B07_aoflagger_polcal_flag_status False
preflag_B07_fluxcal_shadow False
preflag_B07_polcal_manualflag_baseline ['']
preflag_B07_aoflagger_bandpass_status True
preflag_B07_polcal_edges False
preflag_B07_fluxcal_manualflag_clipzeros False
preflag_B07_polcal_manualflag_antenna ['']
preflag_B07_fluxcal_manualflag_corr ['']
preflag_B07_fluxcal_manualflag_channel ['']
preflag_B07_targetbeams_manualflag_time ['']
preflag_B07_targetbeams_manualflag_from_file False
preflag_B07_polcal_manualflag_auto False
preflag_B07_aoflagger_fluxcal_flag_status False
preflag_B07_polcal_ghosts False
preflag_B07_targetbeams_manualflag_corr ['']
preflag_B07_targetbeams_manualflag_antenna ['']
preflag_B07_polcal_manualflag_clipzeros False
preflag_B07_targetbeams_manualflag_channel ['0~1024']
preflag_B07_polcal_manualflag_channel ['']
preflag_B07_polcal_manualflag_time ['']
preflag_B07_targetbeams_shadow True
preflag_B07_fluxcal_manualflag_from_file False
preflag_B07_polcal_shadow False
preflag_B07_targetbeams_manualflag_baseline ['']
preflag_B07_aoflagger_targetbeams_flag_status True
preflag_B07_polcal_manualflag_corr ['']
preflag_B07_targetbeams_ghosts False
	\end{verbatim}
	\vdots
	
		\section{Changes to the MIRIAD source code before compilation}
	\label{appendix_miriad}
	
	\begin{verbatim}
gpcopy.for: l.469  if (relax) int1 = max(int1, 1.0d0)
fits.for: l.454   call uvset(tno,'corr','r',0,0.0,0.0,0.0)
          l.461   call uvset(tno,'preamble','uv/time/baseline',0,0.0,0.0,0.0)
          l.463   call uvset(tno,'preamble','uvw/time/baseline',0,0.0,0.0,0.0)
mfcal.h   l.4     parameter(MAXDATA=1000000,MAXPOL=2)
uvio.c    l.3549  if(u*u + v*v <= limit && w <= 0) return(1);
maxdim.h         PARAMETER(MAXANT = 16)
                 PARAMETER(MAXCHAN = 32768)
                 PARAMETER(MAXWIN = 32)
                 PARAMETER(MAXWIDE = 32)
                 PARAMETER(MAXFBIN = 32)
maxdimc.h        #define MAXANT  16
                 #define MAXCHAN 32768
                 #define MAXWIN 32
                 #define MAXWIDE 32
	\end{verbatim}
	\end{appendix}
	
\twocolumn

	\bibliography{bibtex}{}
	\bibliographystyle{model2-names}

\end{document}